\documentclass[12pt]{article}

\input epsf.sty
\newcommand{\nl}{\hspace{-.65cm}}
\newcommand{\be}{\begin{equation}}
\newcommand{\ee}{\end{equation}}
\newcommand{\ben}{\begin{eqnarray}\displaystyle}
\newcommand{\een}{\end{eqnarray}}

\newcommand{\sectiono}[1]{\section{#1}\setcounter{equation}{0}}

\def\sqr#1#2{{\vcenter{\vbox{\hrule height.#2pt
         \hbox{\vrule width.#2pt height#1pt \kern#1pt
            \vrule width.#2pt}
         \hrule height.#2pt}}}}

\def\ifig#1#2#3{\xdef#1{fig.~\the\figno}
\writedef{#1\leftbracket fig.\noexpand~\the\figno}%
\global\advance\figno by1}

\def\sqr#1#2{{\vcenter{\vbox{\hrule height.#2pt
         \hbox{\vrule width.#2pt height#1pt \kern#1pt
            \vrule width.#2pt}
         \hrule height.#2pt}}}}

\begin{document}

{}~ \hfill\vbox{
%\hbox{hep-th/yymmnnn}
  }\break

\vskip 1cm

\begin{center}
\large{\bf Inflection Point Inflation and Time \\Dependent
Potentials in String Theory}

\vspace{10mm}

\normalsize{Nissan Itzhaki and Ely D. Kovetz\\\vspace{2mm}{\em
Tel-Aviv University, Ramat-Aviv, 69978,
Israel}\\\vspace{2mm}nitzhaki@post.tau.ac.il,~elykovetz@gmail.com}

%\vspace{10mm}

\end{center}\vspace{10mm}

\begin{abstract}

\medskip

We consider models of inflection point inflation.  The main drawback
of such models  is that they suffer from the overshoot problem.
Namely the initial condition should be fine tuned to be near the
inflection point for the universe to inflate.  We show that stringy
realizations of inflection point inflation are common and offer a
natural resolution to the overshoot problem.

\end{abstract}

\newpage

\baselineskip=18pt

\sectiono{Introduction }

Recent experimental data \cite{wmap3} provides evidence for an early
universe inflation \cite{oi,ni0,ni} and quite remarkably even makes
predictions about some of the parameters that characterize models
of inflation.

String theory, however, does not seem to provide, as yet,  a
particularly natural setup for inflation (for reviews of recent
progress see \cite{linde,kallosh}). The basic issue is that in order
to generate a large amount of inflation, typically, the expectation
value of the inflaton needs to vary over super-Planckian distances
which is not easy to achieve in string theory where the inflaton
usually has a geometrical meaning.
%In an effective field theory setup this does not necessarily suggest ...
%,but in string theory the expectation values of the fields often has a geometrical meaning
%that implies a cutoff on the allowed distances.

Recently
\cite{talk,Baumann:2007np,Baumann:2007ah,Krause:2007jk,Panda:2007ie}
it was realized that there is  a relatively simple way to evade the
super-Planckian problem in string theory (for earlier work in the
context of MSSM see \cite{Allahverdi:2006,Bueno
Sanchez:2006xk,Allahverdi:2007}). If the inflaton potential has an
inflection point (or an almost inflection region) then a large
amount of inflation will be generated around the inflection point
(region), provided that the inflaton spends enough time in this
region. Thus if the initial condition for the inflaton is near the
inflection point, then the number of e-foldings will be large (in
fact, very large). However, if the initial condition of the inflaton
is away from the inflection point (which is the generic case), then
the inflaton will overshoot the inflection point without inflating
the universe. In other words, the Hubble friction is not sufficient
to slow down the inflaton at the inflection point, and the inflaton
will not spend enough time near the inflection point to generate
inflation. This can be viewed as a limit of the overshoot problem
discussed in \cite{rami}.

The aim of this paper is twofold. First, we demonstrate in the
context of modular inflation \cite{mi1,mi2} that there are a lot of
simple examples of inflection point potentials in string
theory.\footnote{More complicated example, in the context of brane
inflation were discussed recently in
\cite{Baumann:2007np,Baumann:2007ah,Panda:2007ie,Krause:2007jk}.}
Second, we show that quite generically string theory resolves the
main drawback of inflection point inflation - the overshoot problem
discussed above. The nice aspect of this resolution is that it
involves stringy degrees of freedom and so the supergravity fields
by themselves are not sufficient (for other examples of stringy
degrees of freedom that help in stabilization of moduli fields,
namely through trapping by particle production near ESPs, see
\cite{Kofman:2004yc,Watson:2004aq,Cremonini:2006sx,Greene:2007sa}).

The paper is organized as follows. In section 2 we summarize some of
the general features of inflection point inflation that are relevant
for our work. In section 3 we review the different terms in the
potential for the radion and show that an inflection point potential
can be easily constructed using these terms. We start section 4 by
numerically demonstrating the overshoot problem associated with
inflection point inflation. Then we describe the stringy mechanism
that resolves this problem. We conclude with some comments in
section 5.

\sectiono{Inflection point inflation}

In this section we collect  some of the general properties of
 inflection
point inflation (IPI). Most of the observations made in this section
can be found in \cite{Allahverdi:2006,Bueno
Sanchez:2006xk,Allahverdi:2007}.

We assume that the inflaton, denoted by $\phi$, has a canonically
normalized kinetic term and a potential with an inflection point at
$\phi_{inflection}$. Therefore the potential near
$\phi_{inflection}$ takes the form
 \be\label{45} V(\phi)= V_{inflection}-\beta (\phi-\phi_{inflection})^3 , \ee
and  the slow roll parameters are
 \be \epsilon =\frac12 \left( \frac{V^{'}}{V}\right)^2=\frac{9\beta^2 (\phi-\phi_{inflection})^4}{2V_{inflection}^2},~~~~~~~
 \eta=\frac{V^{''}}{V}=-\frac{6\beta (\phi-\phi_{inflection})}{V_{inflection}}. \ee
We see that near $\phi_{inflection}$ the conditions for slow roll
inflation are satisfied, and that $\epsilon\ll \eta$. This will play
a role momentarily.

Theoretically the nicest feature  of IPI is that during inflation
the expectation value of $\phi$ need not vary over super-Planckian
distances as can be seen from
 \be\label{n}
N=\int\frac{V}{V^{'}} d\phi \approx \frac{V_{inflection}}{3\beta
 (\phi_{start}-\phi_{inflection})}.\ee
This is of particular importance in string theory where the
inflaton typically has either a direct or indirect (U-dual)
geometrical meaning, which makes it problematic to vary it over
super-Planckian distances.

At a more practical level the big advantage of IPI is that if the
low energy approximation one uses is valid at $\phi_{inflection}$
then it is valid throughout the period of inflation. In the context
of string theory this means that if the supergravity approximation
is valid at $\phi_{inflection}$ then it can be used to describe the
whole process of inflation (but not necessarily of re-heating). This
is in contrast with other models of inflation where, typically, the need to
generate a large number of e-foldings pushes the
inflaton either away from the region of validity of the
approximation or away from the region of slow roll.

 As far as observation goes the sharpest prediction of
IPI is that the spectral index, $n_s$, is smaller than $1$ by a
considerable amount. This is obtained from the following
consideration. In the slow roll approximation the spectral index is
given by
 \be n_s\approx1-6\epsilon+2\eta\approx 1+2\eta\approx1- \frac{12 \beta
 (\phi_{start}-\phi_{inflection})}{V_{inflection}},\ee
where we have used the fact that $\epsilon\ll \eta$. Combining this
with (\ref{n}) we find that
 \be\label{7} n_s\approx 1-\frac{4}{N} \approx 0.933,\ee
where the last estimate is for $N=60$. This value of $n_s$ is within
the range of current observational limits that give (for small $r$)
$n_s=0.95\pm0.02$.

It would have been great  if  eq.(\ref{7}) was  a sharp prediction
of IPI. This, however, is not the case since there are at least two
kinds of corrections to eq.(\ref{7}) that are not negligible. The
first is due to the fact that a more generic situation is to have an
approximate inflection point than an exact inflection point. This
will clearly modify the prediction for $n_s$. The second
modification is due to time dependent potentials that appear in
string theory and will play a key role in the present paper. Thus
eq.(\ref{7}) should not be viewed as an exact prediction of the
models discussed here. However, since both modifications are
expected to be small, the fact that $n_s$ is smaller than $1$ by a
significant amount is a prediction of the models of (almost) IPI.

As mentioned above a key feature of IPI is that the universe
inflates only when $\phi \sim\phi_{inflection}$. This implies that
models of  IPI are highly sensitive to the initial condition. If the
initial condition is such that $\phi_{initial}$ is near
$\phi_{inflection}$ then the model works fine, in the sense that $N$
is large. But for generic values of $\phi_{initial}$ the inflaton
acquires a large velocity by the time it reaches $\phi_{inflection}$
and it  simply crosses the inflection point without inflating the
universe. Simply put, the inflaton overshoots the inflection point.
It is the purpose of this paper to show that stringy realizations of
IPI resolve this problem.

For later use we also mention that the COBE normalization condition
 gives
 \be\label{cobe} \frac{V_0}{\beta^2} \approx 0.33 ~ 10^{8}  N^4.\ee
In the next section we shall see that it is easy to satisfy this
condition in the stringy models of inflection point inflation.

\sectiono{IPI in string theory}

Our goal in this section is to show that models of IPI can be found
in string theory. In the context of brane inflation this was shown
in \cite{Baumann:2007np,Baumann:2007ah,Krause:2007jk}. Here we focus
on modular inflation in which one of the moduli fields is the
inflaton and the other moduli fields are assumed to be stabilized
with a higher mass than the characteristic mass scale associated
with the inflaton (for a recent review on moduli stabilization see
\cite{dk}). We consider the case where the inflaton is the radion.
We start by summarizing the various known contributions to the
potential of the radion, and then we show that they can  be combined
to yield IPI.

The setup we work with is the usual one with a compact manifold
$M$ of dimension $d=6$. For simplicity we assume that the compact
manifold is characterized by one length scale, $L$, and so the
volume of $M$ is $\mbox{Vol}_M=c L^d$, where $c$ is a dimensionless
constant of order one.

Before we discuss the various ways to generate a potential for $L$
we recall that it is useful to describe the potential in the
Einstein frame, where the kinetic term for $L$ does not mix with the
 kinetic term of the graviton. This ensures that the potential we
find for $L$ can be interpreted in  the standard way. We start with
the Einstein-Hilbert action in $s=4+d$ dimensions and KK reduce it
to four dimensions to find
 \be\label{gt} \frac{1}{16\pi G_s}\int d^s x \sqrt{g_s} R_s,~~~~\Rightarrow~~~
 \frac{c}{16\pi G_s}\int d^4 x \sqrt{g_4} R_4 L^d. \ee
We see that in this frame the four dimensional  Newton constant is
$G_s/(cL^d)$, and so it depends on the expectation value of the four
dimensional field $L$. To suppress this dependence we rescale the
metric
 \be\label{rescale} g_{4_{\mu\nu}} \rightarrow g_{E_{\mu\nu}}= (L/L_0)^d g_{4_{\mu\nu}}, \ee
where $L_0$ is a constant. This leads to the standard 4d EH action
 \be S_1=\frac{c L_0^d}{16\pi G_s}\int d^4x \sqrt{g_E} R_E,\ee
where the 4D Newton constant does not depend on the expectation
value of $L$
 \be G_N=\frac{G_s}{cL_0^d}.\ee

Now that the kinetic terms of $L$ and the graviton do not mix we can
turn to the various static contributions to $V(L)$.

\nl $\bullet$ Contributions due to (3+p)-branes:\\ By (3+p)-branes
we mean branes that wrap $p$ cycles in $M$  and all the four
non-compact directions. Namely, from a four dimensional perspective
these are space-filling branes. Their action in the frame (\ref{gt})
is
 \be\label{98} N_{p+3} T_{3+p} \int d^4 x \sqrt{g_4} L^p,\ee
 where $N_{p+3}$ is the number of (3+p)-branes and $ T_{3+p}$ is their tension.
 To find the
potential  in the Einstein frame we rescale (\ref{rescale}). This
rescaling gives a factor of $( L_0/L)^{d/2}$ for each of the four
non-compact directions. Therefore the potential is
 \be\label{3} V_{3+p}=N_{3+p} T_{3+p}L_0^{2d}
 \frac{1}{L^{2d-p}}.\ee
It is interesting to note that  this potential  goes to zero in the
decompactification limit ($L\to\infty$). This is a somewhat
counterintuitive result since in this limit the size of the brane
blows up like  $L^p$ and so does the action (\ref{98}). It is only due
to the rescaling to the Einstein frame that  the potential vanishes.

\nl $\bullet$ Contributions due to l-fluxes:\\  The relevant term in
the  $4+d$ dimensional action is
 \be \int d^s x \sqrt{g_s} c_l F_l^2 ,\ee where the $c_l$'s are
 constants that typically depend on the value of the other moduli.
Now suppose that we have $N$ units of $F_l$ (i.e. $\int_{l-cycle}
F_l=N_l$) then we find in the 4d effective action (in the Einstein
frame) the following potential for $L$
 \be V_{l-flux}=N_l^2 c_l L_0^{2d}  \frac{1}{L^{d+2l}}.\ee

\nl $\bullet$ Contributions due to the curvature of $M$:\\ The EH
action in $4+d$ leads also to the following term in 4D
 \be \frac{1}{G_s} \int d^4 x \sqrt{g_4}(L_0/L)^{2d}(\int d^dx\sqrt{g_d} R_d). \ee
On dimensional ground $\int d^dx\sqrt{g_d} R_d=k_M L^{d-2}$ where
$k_M$ is a dimensionless parameter that (depending on the topology
of $M$) can be negative zero or positive. Thus we find
 \be V_{cur}=\frac{k_M L_0^{2d}}{G_s} \frac{1}{L^{d+2}}.\ee
Note that when $M$ is  a CY manifold $k_M$ vanishes.

There are also non-perturbative contributions due to  (-1+p)-branes
and gaugino condensation. These are exponentially small in $L$ and
will not play a role here.

When attempting to use these potentials to construct models of
inflation it is important to make sure that the scalar is
canonically normalized. This ensures that also in dynamical
situations, such as inflation, the potential has the usual
interpretation.  In the case at hand we denote the canonically
normalized scalar field associated with $L$ by $\phi$ and the two
are related in the following way (for $d=6$)
 \be\label{nor} L= e^{\alpha
 \phi}~~~~\mbox{with}~~~~\alpha=
 %\frac{1}{4\sqrt{3}}.
 \frac{1}{\sqrt{24}}.\ee

Before discussing IPI let us show that none of the terms discussed
above leads to inflation by itself. With  a single contribution
$V(L)$ takes the form
 \be\label{po}
 V(\phi)= \frac{V_0}{L^C}=V_0\exp\left( -\sqrt{\frac2k}
 \phi\right)~~~~~\mbox{with}~~~~~ k=48/C^2.\ee
Such a potential is known to yield exact cosmological solutions
\cite{lm}. For a flat universe
 \be  ds^2=-dt^2+a(t)^2dx_i^2~~~i=1,2,3\ee
we have
 \ben\label{tye} && a(t)=a_0 t^k,\\ \nonumber
 && \phi=\sqrt{2k}\log\left( \sqrt{\frac{V_0}{k(3k-1)}} t\right) , \een
which implies a universe with
 \be w=\frac{P}{\rho}=-1+\frac{2}{3k},\ee
where as usual $P$ is the pressure and $\rho$ is the energy density.
Since all the examples mentioned above have $C\geq 7$ and since the
condition for an accelerating universe is $k>1$ we see from
(\ref{po}) that none of the stringy contributions to $V(L)$ give an
accelerating universe, let alone a universe with $n_s$ close to
$1$.\footnote{Moreover even if there was a stringy potential with a
small enough $C$ to accelerate the universe  there would still be
the problem of a graceful exit. Namely, with such a potential
inflation will not end. This is, of course, interesting for models
of quintessence but not for inflation.} This is believed
\cite{Hellerman:2001yi} to be a general result valid for all moduli,
not necessarily the radion.

It is therefore particularly interesting that a combination of
these potentials can give an inflection point inflation. For this to
happen  we need the potential to get contributions from three terms
 \be\label{static} V=a_1 \exp(j_1\alpha \phi) +a_2\exp(j_2\alpha \phi)+a_3 \exp(j_3\alpha
 \phi),\ee
with, say, $j_1>j_2>j_3$, $a_1, a_3 >0$ and $a_2<0$.

Note that the requirement  $a_2<0$ can be satisfied since not all
the terms discussed above are  positive definite. $V_{cur}$, for
example, can  be negative. When $L$ is small this might lead to some
non-trivial  effects \cite{Silverstein:2005qf,McGreevy:2006hk} due
to winding modes that become light. We, however,  are interested in
the large $L$ case in which nothing dramatic is expected to happen
if $V_{cur}$ is negative. Another way to generate  a negative
potential is via orientifolds. Their tension  is negative and so
they induce negative $V_{3+p}$. This puts some mild constraints on
the possible values of $N_{3+p}$ that can  be satisfied. To
conclude, using $V_{p+3}, ~V_{l-flux}$ and $V_{cur}$ one can
construct many examples of inflection point inflation in string
theory.

For concreteness in the rest of the paper we focus on a particular
example with $j_1=12$, $j_2=10$ and $j_3=8$. We emphasize, however,
that the conclusions we present below do {\it not} depend on this
particular choice. In particular, the stringy resolution of the
overshoot problem works equally well for other choices of $j_1$, $j_2$
and $j_3$.

For $j_1=12$, $j_2=10$ and $j_3=8$ an inflection point at
$\phi=\phi_{inflection}= \sqrt{24}\log(L_{inflection})$ is obtained
if we take
 \be\label{qq} \frac{a_1}{a_3}=\frac23 ~L_{inflection}^4,
 ~~~~~\frac{a_2}{a_3}= -\frac85  ~L_{inflection}^2.\ee
Since an overall rescalling of the potential does not change
$\phi_{inflection}$ only ratios of the $a$'s appear in this
condition. From (\ref{qq}) we see that, as expected,  to have a
potential with an inflection point we need to fine tune the
parameters in the potential
 \be\label{finetune} \frac{a_2^2}{a_1 a_3}=\frac{96}{25}.\ee
In the last section we shall see that this condition can be relaxed.

Expanding near the inflection point we find that \be
V_0=\frac{a_3}{15 L_{inflection}^8}, ~~~~
\beta=\frac{a_3\sqrt{24}}{54 L_{inflection}^8},\ee
 where we are using  the notation     of eq. (\ref{45}), and so
the COBE normalization condition, (\ref{cobe}), gives
 \be L_{inflection}\approx  6.7 ~a_3^{1/8}\sqrt{N},\ee
which implies that for typical values of $a_3$ the supergravity
approximation is valid. This is not surprising since usually the
COBE normalization condition implies that the energy scale
associated with inflation is a few orders of magnitude smaller than
the Planck scale.

\sectiono{The overshoot problem and its stringy resolution}

As explained above the main drawback of  models of IPI is that they
suffer from the overshoot problem. This problem is generic  and
appears also in the stringy realization of models of IPI
 discussed in the previous section. This is
illustrated in figure 1. In this figure we see how sensitive
inflation is to the initial condition. If the initial condition is
near the inflection point then the number of e-foldings, $N$, is
large. But if the initial condition is even slightly away from the
inflection point, inflation does not occur since $\phi$ overshoots
$\phi_{inflection}$. Interestingly enough string theory provides a
simple dynamical resolution to this problem which we shall now
discuss.

In the previous section we described some of the static terms in
$V(L)$. There are, however, also time dependent contributions to
$V(L)$. These are due to particles with masses that depend on $L$.
Denoting the particle densities by $n_i$, the potential they
induce is
 \be V_{i} (L,t)=n_i(t) m_i(L).\ee
This potential is time-dependent because $n_i(t)$ dilutes as the
universe expands
 \be\label{ok} n_i(t) \sim \frac{1}{a^3(t)}.\ee
Most known examples in string theory, such as perturbative states,
yield $ m(L)\to 0$ in the decompactified limit, and so $V_{i}(L,t)$
 vanishes in this  limit and does not help much with the overshoot problem.
\begin{figure}
\begin{picture}(110,120)(0,0)
\vspace{100mm} \hspace{-61mm} \mbox{\epsfxsize=58mm
\epsfbox{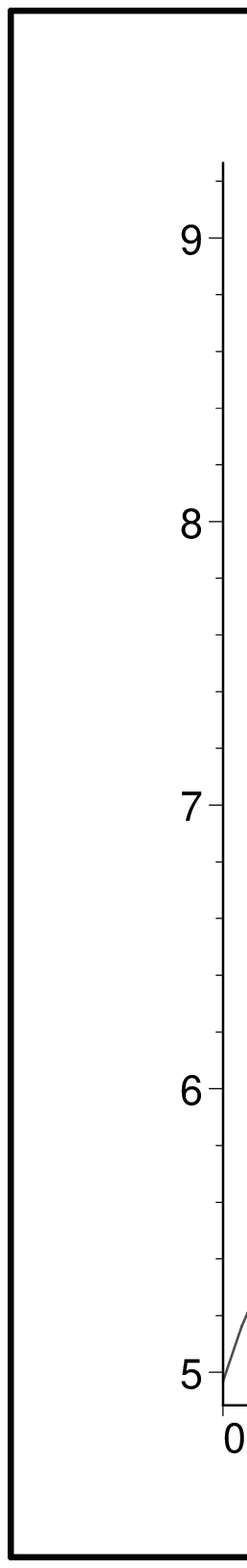}}
\end{picture}
\begin{picture}(110,120)(0,0)
\vspace{0mm} \hspace{-21mm} \mbox{\epsfxsize=58mm
\epsfbox{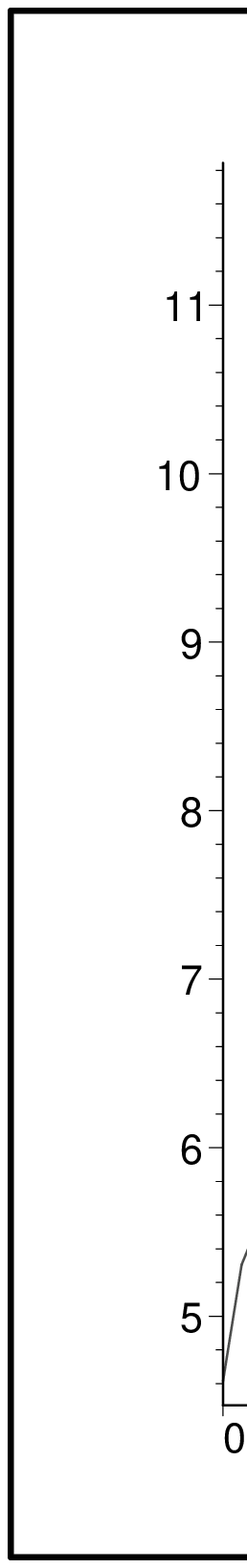}}
\end{picture}
\caption{A numerical illustration of the overshoot problem: In both
cases we take $L_{inflection}=3$ and display $\phi$ as a function of
time. In the picture to the left we take
$\phi_{initial}=\frac{12}{13} \phi_{inflection}$ and obtain a decent
amount of inflation when $\phi$ crosses $\phi_{inflection}\approx
5.38$. In the picture to the right we take
$\phi_{initial}=\frac{6}{7} \phi_{inflection}$. Despite the fact
that this is fairly close to $\phi_{inflection}$ inflation is not
generated. }
\end{figure}
However,   non-perturbatively there are (0+p)-branes (namely, branes
that wrap $p$ cycles in $M$ and are point-like objects in the four
dimensional uncompactified space-time)  with $m(L)\to\infty$ in the
decompactified limit.

To be precise the mass of a (0+p)-brane  is
 \be\label{mass} m_{0+p}=T_{0+p} L^p \left( \frac{L_0}{L}\right) ^{d/2},\ee
where $T_{0+p}$ is the tension of the brane. The factor of $L^p$ is
due to the volume of brane, while the $(L_0/L) ^{d/2}$ factor is due
to the transformation to the Einstein frame (coming from the
$\sqrt{g_{tt}}$ in the particle action). We see that, unlike in the
case of (3+p)-branes (\ref{3}), now, for $p>d/2$, the volume term
dominates and we find that $m(L)\to\infty$ in the decompactification
limit. Therefore, such (0+p)-branes lead to time dependent
potentials that blow up when $L\to\infty$
 \be\label{par} V_{0+p}=n_{0+p} T_{0+p} L_0^3 L^{p-3} .\ee

We would like to argue now that this kind of time dependent
potential is exactly what is needed to resolve the overshoot problem
of IPI. The basic idea is  simple and illustrated in figure 2. The
static piece of the potential is denoted by the red/solid line and
the time dependent piece of the potential by the green/dashed line.
In (a),(b) the static potential is steep, but since there has not
been significant expansion the time-dependant potential is able to
balance it, preventing $\phi(t)$ from acquiring a large velocity.
Heuristically, $\phi(t)$ (denoted by the blue/filled circle) follows
the dilution of the time dependent potential. In (c) $\phi(t)$
enters the shallow region (where the slow roll parameters are small)
where inflation takes place and the time dependent potential starts to slow down exponentially.
In (d) $\phi(t)$ is dominated solely by the static potential but its
velocity is now low enough to allow prolonged inflation.

\begin{figure}
\begin{picture}(110,120)(0,0)
\vspace{100mm} \hspace{0mm} \mbox{\epsfxsize=37mm
\epsfbox{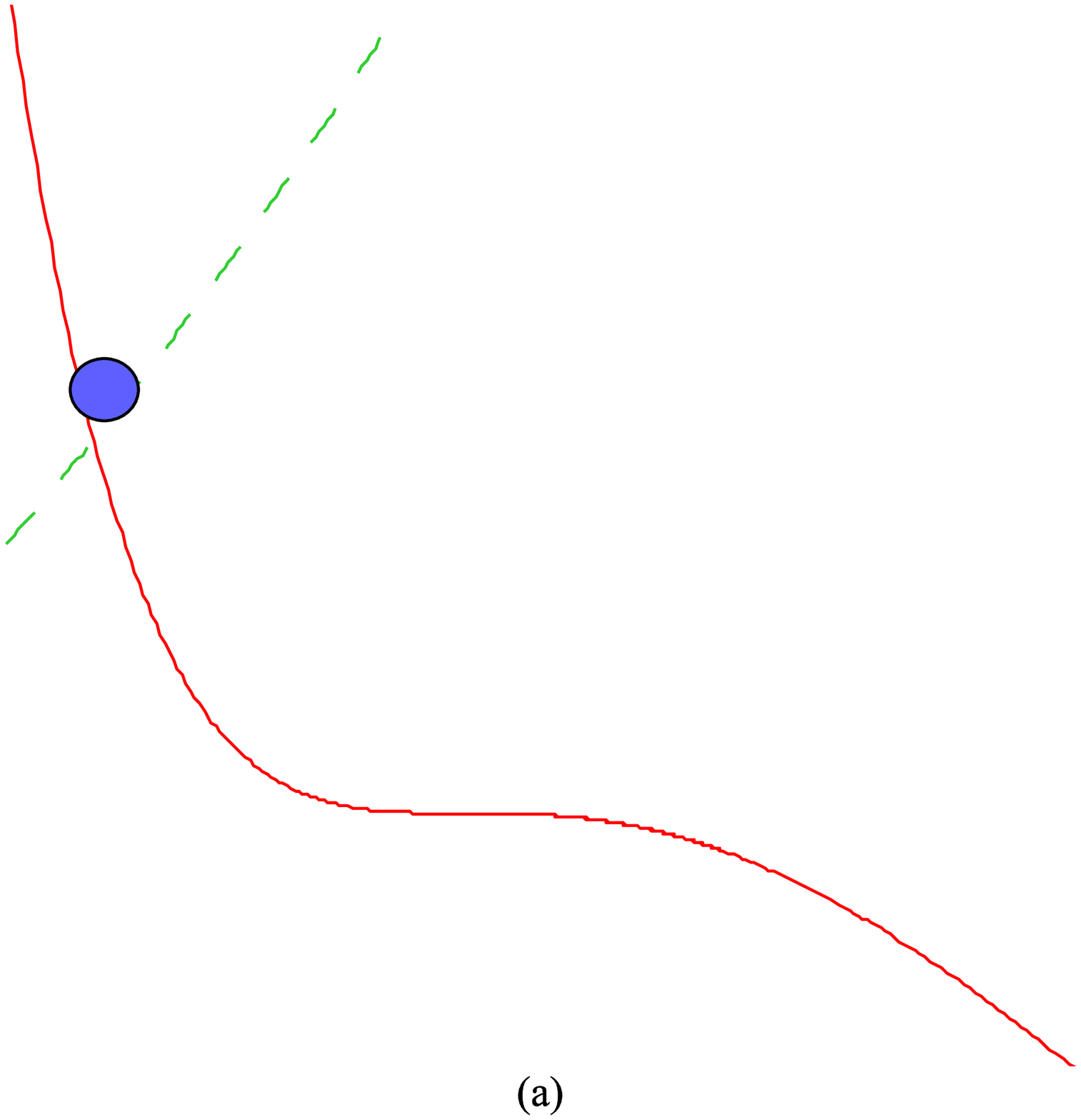}}
\end{picture}
\begin{picture}(110,120)(0,0)
\vspace{0mm} \hspace{0mm} \mbox{\epsfxsize=37mm
\epsfbox{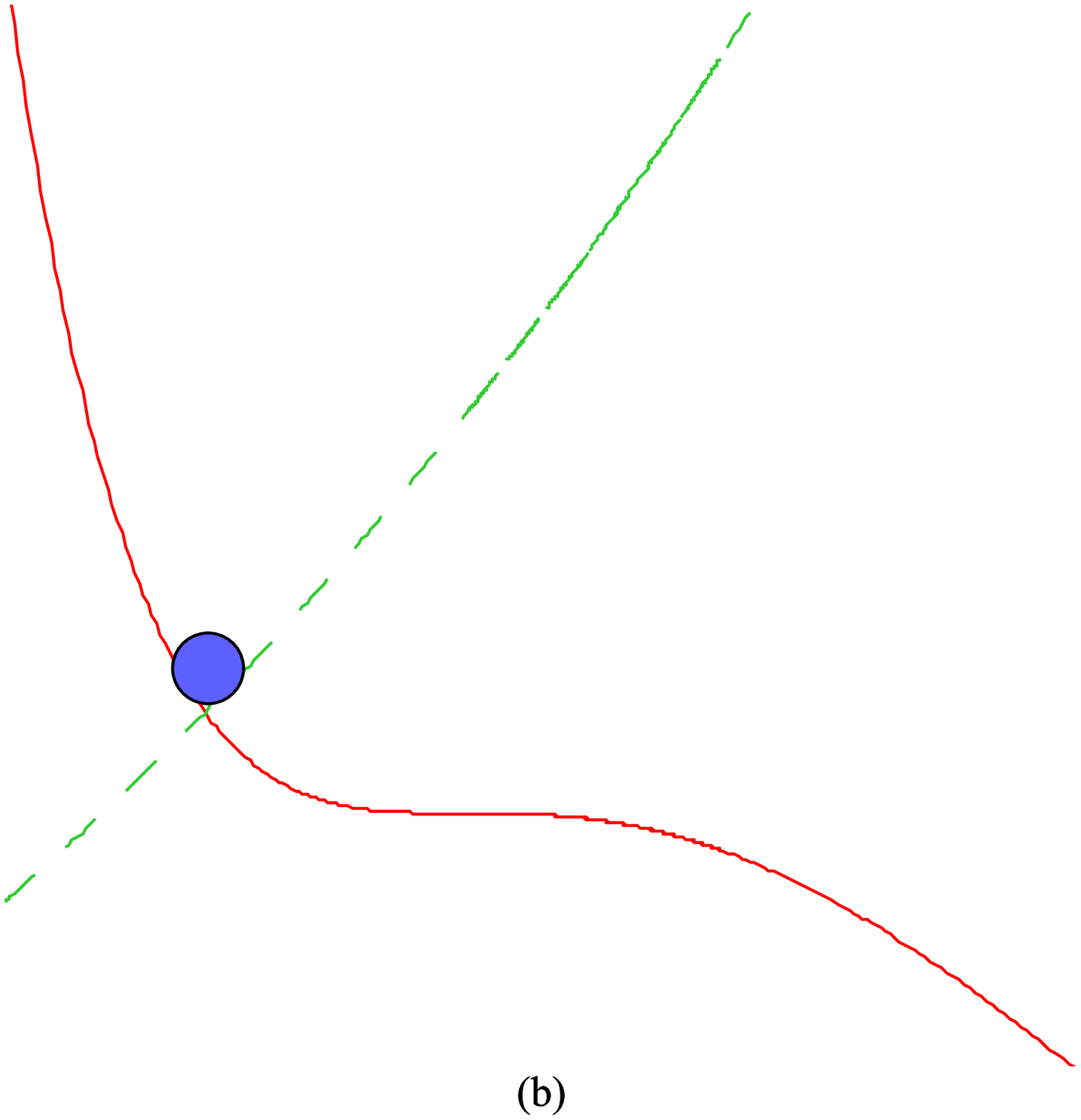}}
\end{picture}
\begin{picture}(110,120)(0,0)
\vspace{0mm} \hspace{0mm}
\mbox{\epsfxsize=37mm\epsfbox{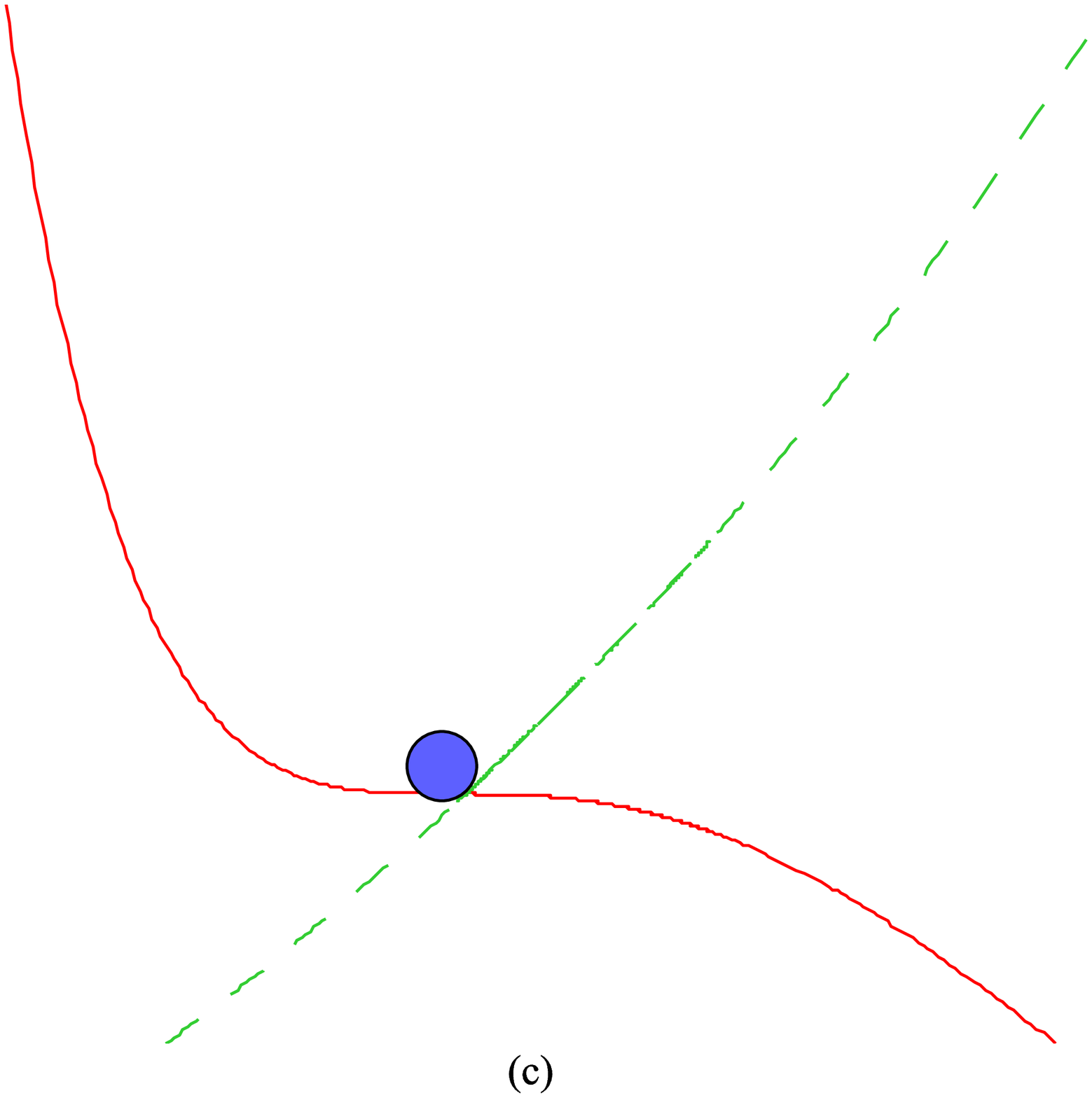} }
\end{picture}
\begin{picture}(110,120)(0,0)
\vspace{10mm} \hspace{0mm} \mbox{\epsfxsize=37mm
\epsfbox{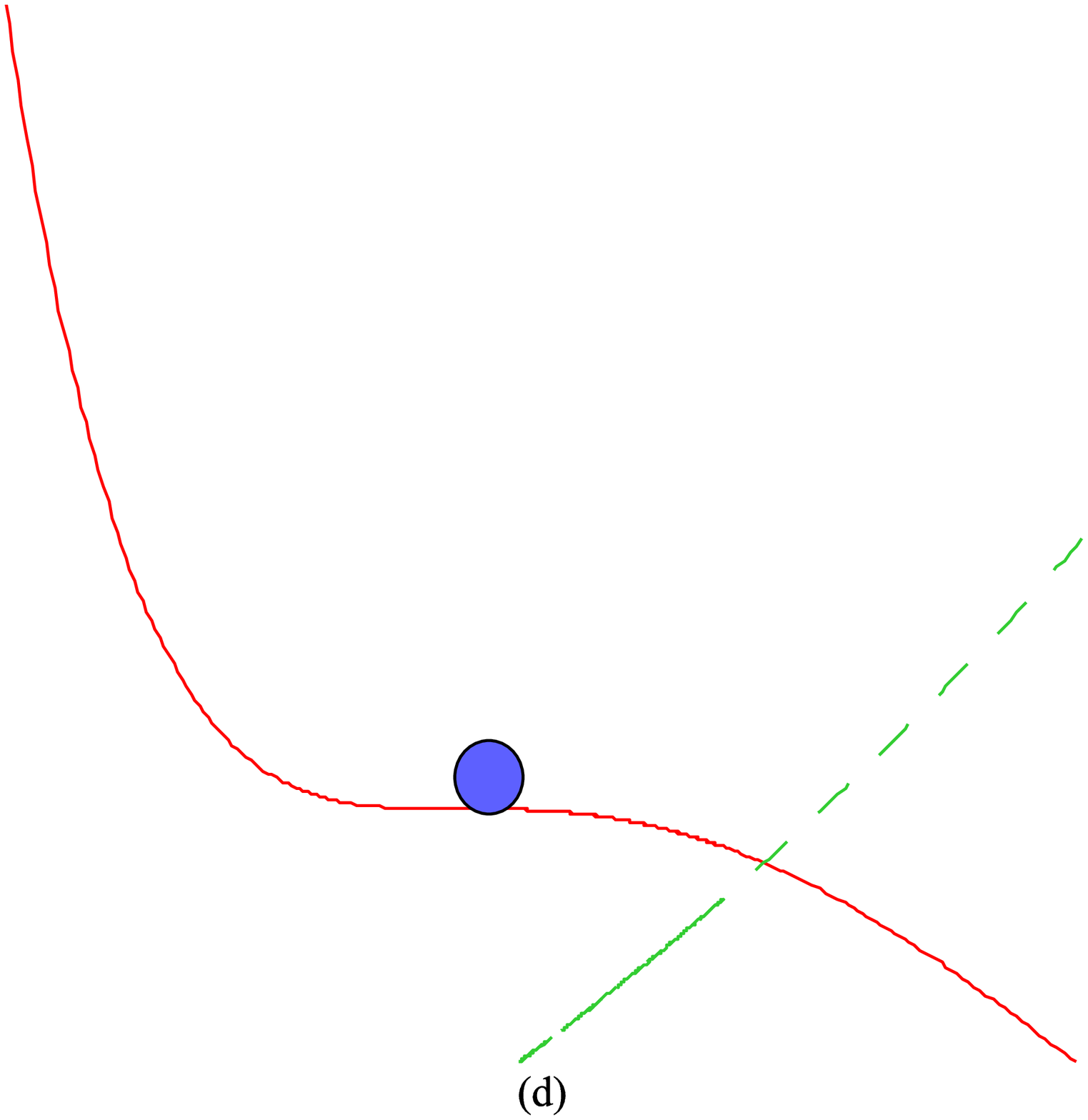}}
\end{picture}
\caption{ A heuristic demonstration of how a time dependent
potential, that scales like $1/a^3(t)$, can resolve the overshoot
problem of IPI.}
\end{figure}

To see if this heuristic argument actually holds we have to solve
the equations of motion which for a flat FRW universe  are (we take
the reduced Planck mass to be $1$)
 \ben\label{itr} 3H^2&=&\frac12 \dot{\phi}^2 + V_{static} +V_{0+p}, \nonumber\\
  \dot n_{0+p}&=&-3 H n_{0+p},\\
  \ddot{\phi}+3 H \dot{\phi}& =&
%  -\acute{V}_{static}-\acute{V}_{0+p},\nonumber
 -\frac{d}{d\phi}\left( V_{static} +V_{0+p}\right)\nonumber
 \een
where $V_{static}$ is given by (\ref{static}, \ref{qq})  with
$j_1=12$, $j_2=10$, $j_3=8$ and $V_{0+p}$ by (\ref{par}). We assume that
other moduli fields, in particular the dilaton, are not too extreme
and so all the constants that appear in $V_{static}$ and $V_{0+p}$
are of order $1$ in string units. We will get back to this point
momentarily.

The question that we wish to address is whether a generic initial
condition will give a large enough $N$ to be consistent with
experiment. Since we are working within the supergravity
approximation, the approximation breaks down at $L<1$. This means
that we should take $L_{inflection}>1$ (for the approximation to be
valid during inflation) and that the most natural initial condition
is $L_{initial}\sim 1$ (which gives $\phi_{initial}\sim 0$). Recall
that the illustration in the beginning of this section shows  (for
$n_{0+p}=0$) that for these kind of initial conditions the inflaton
overshoots the inflection point and the universe does not inflate.
In fact figure 1 shows that this happens already when
$L_{initial}/L_{inflection}=6/7$.

To see what happens for $n_{0+p}>0$ we have to solve (\ref{itr})
numerically. On general grounds we expect $n_{0+p}^{initial-min}$,
which is  the minimal initial value of $n_{0+p}$ (with
$\phi_{initial}= 0$ and $\dot{\phi}_{initial}=0$) that is needed to
generate a significant amount of inflation, to become smaller as we
increase $L_{inflection}.$ The reason is simply that $V_{static}$
becomes weaker at large $L$ while $V_{0+p}$ becomes stronger.

The table below summarizes our findings for $p=6$, $a_3=1$ and
$N=100$
\begin{center}
\renewcommand{\arraystretch}{2}
     \begin{tabular}{|c|c|c|c|c|c|c|c|c|c|c|c|}
       \hline
            $L_{inflection}$ & $2$
             &$5$ &$8$&$10$&$15$&$20$&$25$\\
             \hline
                      $n_{0+6}^{initial-min}$    & $0.098877$ &
                      $0.005431$ & $0.00081$&  $0.000316 $& $0.000054$&$  0.000015 $&$
0.000005$\\
       \hline
                   \end{tabular}
\end{center}
We see that the needed values of $n_{0+p}^{initial}$ are small and
 become smaller and smaller as we increase
$L_{inflection}$. This is particularly important because it implies
that in the supergravity region, $L_{inflection}\gg 1$, a tiny
$n_{0+p}^{initial}$ is sufficient
 to slow down $\phi$ enough by the inflection point for the universe
 to inflate significantly.

The nice feature of this mechanism is that a small
$n_{0+p}^{initial}$ is expected to be  generated by  quantum effects
and need not be imposed by hand! The reason is the following. When
$L_{initial}\sim 1$ the Hawking temperature associated with the
initial vacuum energy is $T_H\sim \sqrt{V_{initial}}\sim1$. Since
the mass of the $(0+p)$ -brane is also of order $1$ we expect
$n_{0+p}^{initial}$ to be of order $1$ as well. Thus
$n_{0+p}^{initial}$ bigger than the values that appear in the table
above is expected to be generated quantum mechanically.

This discussion depends, of course, on the value of the other moduli
that we assumed to be stabilized. In particular the value of the
dilaton has a significant effect since $T_{H}\sim g$ while
$M_{0+p}\sim 1/g$, and so a natural $n_{0+p}^{initial}$ scales like
$e^{-M_{0+p}/T_H}\sim e^{-1/g^2}$. This means that one cannot send
the string coupling constant to zero while fixing $L_{inflection}$.
However, as the table above shows, for a large enough
$L_{inflection}$ we can have $g$ that is considerably smaller than
$1$.

\sectiono{Concluding remarks}

We end with some   comments:\\
$\bullet$ Our numerical simulations indicate that the mechanism
described here works quite generally. For example it works fine also
for $3<p<6$, though it is not as efficient in the sense that
$n_{0+p}^{initial-min}$ is a bit larger than
$n_{0+6}^{initial-min}.$ As mentioned above it also works for other
choices of $j_1$, $j_2$ and $j_3$.\\
$\bullet$ Here we focused on the case where the radion is the
inflaton and show that models of IPI are common in string theory
and resolve the overshoot problem. It should be interesting to see
if this can be generalized to other setups of IPI. A particularly
interesting one is the one  of \cite{Baumann:2007np,Baumann:2007ah,Krause:2007jk}.\\
$\bullet$  Taking $n_{0+6}^{initial}$ to be  bigger than
$n_{0+6}^{initial-min}$ by a factor of order $1$ leads to $N$ that
is practically infinite. This implies that a large enough $N$ is
generated with rather generic initial conditions even when the
potential does not have an exact inflection point but rather an
approximated inflection region. Namely, eq. (\ref{finetune}) can be
relaxed. Thus  one needs to tune but not fine tune the parameters in
the potentials. This also means  that stringy corrections to the
potential are unlikely to change our main conclusions. The fact that
eq. (\ref{finetune}) can be relaxed is  particularly important since
$a_1$, $a_2$ and $a_3$ are quantized once all other moduli are fixed
(which is our working assumption).

To demonstrate this behavior, we explore the range of deviation
from the exact inflection point parameters
 in our example above that would still allow enough e-foldings
 ($N>60$) with $n_{0+6}^{initial}=\frac{1}{100}$. The table below shows
 the maximal deviation for $a_1$ per value of $L_{inflection}$,
  according to $a_1=a_1+\delta{a_1}$ (we allow $a_1$ to vary in
  the direction that causes $V$ to lose its extremal point and monotonically approach zero): \\

\begin{center}
\renewcommand{\arraystretch}{2}
     \begin{tabular}{|c|c|c|c|c|c|c|}
       \hline
            $L_{inflection}$ & $5$ & $8$ & $10$ &$15$&$20$&$25$\\
             \hline
                      $\delta{a_1}$    & $0.01$ & $0.08$ & $0.2$ & $1$ & $3$ & $8$\\
       \hline
              %  $\delta{a_2}$    & $0.0004$ &$0.0010$ & $0.0016$&$0.0036$&$0.0064$&$0.01$ \\
      % \hline
                   \end{tabular}
\end{center}
We see that as we increase $L_{inflection}$ the sensitivity to the
parameters of the potential decreases, and in particular since
$\delta{a_1}$ is of order $1$ the quantization conditions of the
$a_i$'s can be satisfied.

\hspace{5mm}

\noindent {\bf Acknowledgements}

We thank O. Aharony, M. Berkooz, R. Brustein, E. Kiritsis and G.
Veneziano for discussions and especially Y. Oz and S. Yankielowicz
for collaboration at an early stage of the work. This work is
supported in part by the Mary Curie Actions under grant
MIRG-CT-2007-046423.

\end{document}